\documentclass[12pt,reprint]{iopart}
\usepackage{amsmath}
\usepackage{graphicx}
\usepackage{xcolor}
\usepackage{soul}

\begin{document}

\title[Controlling magnon-photon coupling in a planar geometry]{Controlling magnon-photon coupling in a planar geometry}
\author{Dinesh Wagle, Anish Rai, Mojtaba T. Kaffash and M. Benjamin Jungfleisch}

\address{Department of Physics and Astronomy, University of Delaware, Newark, DE 19716, USA}

\ead{mbj@udel.edu}
\vspace{10pt}
\begin{indented}
\item[]January 2024
\end{indented}

\begin{abstract}
The tunability of magnons enables their interaction with various other quantum excitations, including photons, paving the route for novel hybrid quantum systems. Here, we study magnon-photon coupling using a high-quality factor split-ring resonator and single-crystal yttrium iron garnet (YIG) spheres at room temperature. We investigate the dependence of the coupling strength on the size of the sphere and find that the coupling is  stronger for spheres with a larger diameter as predicted by  theory. Furthermore, we demonstrate strong magnon-photon coupling by varying the position of the YIG sphere within the resonator. Our experimental results reveal the expected correlation between the coupling strength and the rf magnetic field. These findings demonstrate the control of coherent magnon-photon coupling through the theoretically predicted square-root dependence on the spin density in the ferromagnetic medium and the magnetic dipolar interaction in a planar resonator.
\end{abstract}

\maketitle


\section{\label{sec:intro} Introduction}

Systems with strong light-matter interaction play an important role in quantum information, communications, and sensing applications such as the quantum internet \cite{kimble08}, quantum memory \cite{kurizki15}, quantum transduction \cite{blum15}, and hybrid quantum devices \cite{wall09}. 
Magnetic materials are ideal candidates for achieving control of strong light-matter interaction because they can have spin densities many orders of magnitude higher than that of dilute spin ensembles \cite{dany19}. Moreover, magnetic media can easily be controlled by external stimuli such as magnetic fields. In  particular, the quanta of collective spin excitations in magnetic materials (i.e., magnons) can interact with  {microwave} photons through light-matter interaction, leading to magnon polaritons~\cite{Mills_Report_74} which host a wealth of interesting physics \cite{soykal10,heubl13,zhang14, tabu14, goryachev_pra_14, bai_prl_15,kli16, bhoi14, bhoi17, kaffash_qst_23}. Magnon-photon coupling relies on the dipolar (Zeeman) interaction between the spins and the magnetic component of the electromagnetic wave of the photon modes. Phenomenologically, coherent magnon-photon coupling can be described using the coupled harmonic oscillator model given by the equation:
\begin{equation}
\label{MPfit}
    f_{\pm} = \frac{1}{2}\left[\left(f_\mathrm{0}+f_\mathrm{r}\right)\pm \sqrt{\Big(f_\mathrm{0}-f_\mathrm{r}\Big)^2+\Big(\frac{g}{\pi}\Big)^2}\right],
\end{equation}
where $f_{\pm}$, $f_\mathrm{0}$ (independent of the applied field), and $f_\mathrm{r}$ (dependent on the applied field) are the hybridized modes, uncoupled resonance frequency of the photon mode, and the ferromagnetic resonance (FMR) mode, respectively. The coupling strength $g/{2\pi}$ governs the extent of coupling, which is larger for systems with a larger number of spins ($N$): $g/{2\pi}\propto\sqrt{N}$~\cite{dicke54}.

\begin{figure}[b]
\centering
\includegraphics[width=0.9\columnwidth]{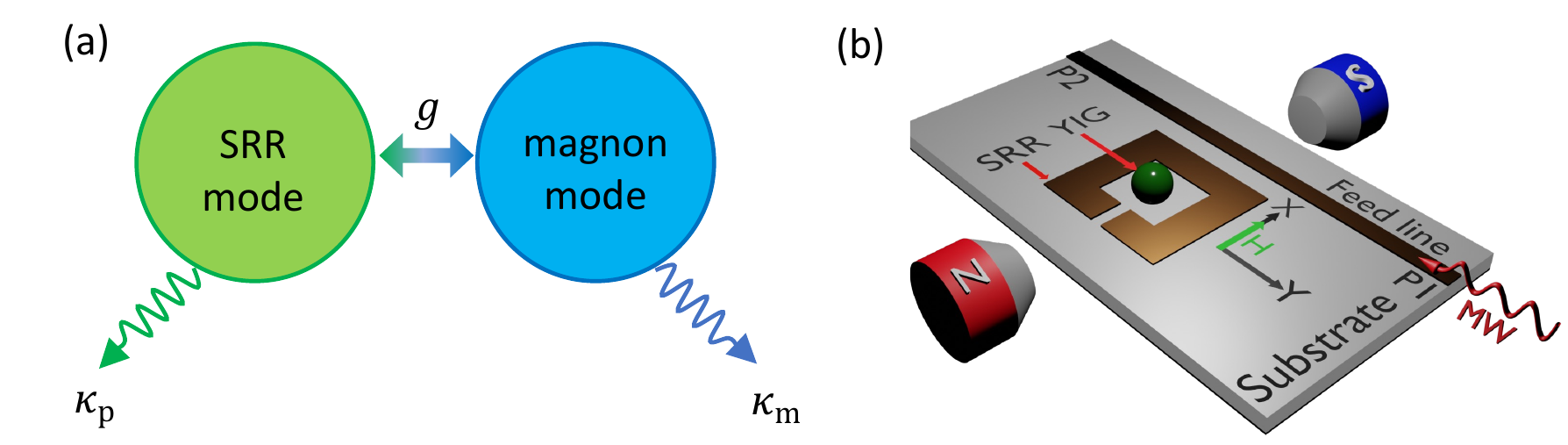}
\caption{(a) Schematic illustration of coupling process between SRR mode (photon) and FMR mode (magnon). $\kappa_p$ and $\kappa_m$ are the microwave photon and magnon dissipation rates, respectively, and $g~(=2\pi g_\mathrm{eff})$ is the magnon-photon coupling strength. (b) Representation of the experimental setup. The resonator consists of a split-ring in the vicinity of the microwave feedline. A YIG sphere is placed on the top of SRR. The experiments are conducted by recording the microwave absorption as a function of the magnetic field.
}
\label{schematic}
\end{figure}

The efficiency of energy exchange between the subsystems of a  {magnonic} hybrid system depends on the strength of coherent magnon-photon coupling. Controlled magnon-photon coupling has previously been demonstrated experimentally in 3D rectangular cavities of different sizes using yttrium iron garnet (YIG) spheres with varying diameters~\cite{zhang14, tabu14} and in a 3D cylindrical cavity by varying the angle between the external magnetic field and the microwave field~\cite{bai_IEEE_16}. Similar physics has been demonstrated in 
 { planar-geometry resonator-based systems, which exhibit higher coupling strengths than those of the 3D cavity resonator-based systems~\cite{Gregory_PRA_15, Kaur_APL_16, Zhang_JPD_17, Girich_SC_23, Gregory_NJP_14},} by varying the position of a YIG thin film~\cite{castel_ieee_17} {. Furthermore, it was shown that the coupling can be controlled in a split-ring resonator with a nonuniform microwave magnetic field by} changing the track width of the split-ring~\cite{shi_JPD_19}. Recently, magnon-photon coupling was shown to be controllable in a NiFe/Pt-superconducting cavity hybrid system by changing the dc current applied to the Py/Pt bilayer~\cite{zhao_APL_23} and in a YIG-superconducting cavity hybrid system by varying the temperature~\cite{zhao_JAP_23}. 
 {Moreover, the ultrastrong coupling regime was reached in superconductor/ferromagnet multilayered heterostructures~\cite{Golov_SA_21}.}
The control of such coupling by spin torque was recently theoretically demonstrated~\cite{Proskurin_PRB_2021,rai_JMMM_23}. 
Furthermore, it was shown that a distinct magnon-photon dissipative coupling regime can be achieved, in which a level attraction can be observed~\cite{harder_PRL_18,bhoi_prb_19}. More recently, the suppression of coherent magnon-photon coupling due to non-linear spin wave interactions at high microwave power was shown \cite{Lee_PRL_2023}. Additionally, the coupling of magnon excitation with a superconducting qubit \cite{Yutaka_Science_2015} and the magnon-photon coupling on a coplanar superconducting resonator~\cite{li19, Ghirri_PRA_23, Li_PRL_22} were demonstrated. 

 {Magnons are bosons, which can couple with other excitations such as photons~\cite{Artman_PR_53, bai15}, phonons {~\cite{Berk_NC_19, Wang_PCCP_23}}, and magnons {~\cite{Li_PRL_20, Pan_PRR_23}} or simultaneously couple to both photons and phonons forming a tripartite hybrid system~\cite{Li_PRL_18, Amazioug_SR_23, Chakraborty_SR_23}. Magnon modes can be controlled using an external magnetic field. On the other hand, photon modes can be manipulated by engineering the resonator design using different geometries and materials. In particular, the photon properties can be altered using periodic structures and Bragg gratings, for example in a photonic crystal 
\cite{Li_Science_18, Narimanov_PRX_14, Foteinoulou_Physics_14}. Ultra-strong magnon-photon coupling has been demonstrated in such photonic crystals with a point defect consisting of a ferromagnetic material~\cite{Zhang_APL_19, Zhang_arXiv_23}.
Furthermore, photonic crystals have been used to enable strong photon-phonon interaction, e.g., \cite{Huang_2003,Zhang_2004,Li_NatureCommunications2014}.
Very recently, hybrid coherent control of magnons in a ferromagnetic phononic resonator has been achieved by laser pulse excitation of a 1D galfenol nanograting~\cite{Scherbakov_PRR_24}}. 
Despite the time and effort invested, efficient control of magnon-photon coupling remains challenging. At the same time, the ability to efficiently control the coupling is essential for realizing on-chip hybrid magnonic devices~\cite{Chumak_Nat_Phys_2015}.

Here, we demonstrate strong magnon-photon coupling in an on-chip planar split-ring resonator (SRR)/YIG sphere system. Our results indicate that coherent magnon-photon coupling can efficiently be controlled by (i) increasing the YIG sphere diameter (corresponding to a higher number of spins)  {upto a critical size for a given resonator}, (ii) choosing the appropriate SRR dimensions and resonance frequency, and/or (iii) increasing the spatial mode overlap between magnon mode in the YIG sphere and the microwave magnetic field.

This article is structured in the following fashion. In Section {2}, we introduce the experimental configuration and setup, in which two SRRs of different resonating frequencies are used to study the position and volume dependence of YIG spheres on the coherent magnon-photon coupling. In Section {3}, we discuss our findings, and in Section 4, we summarize our work.

\section{\label{sec:exp} Experimental configuration and setup}
\begin{figure}[t]
\centering
\includegraphics[width=0.75\columnwidth]{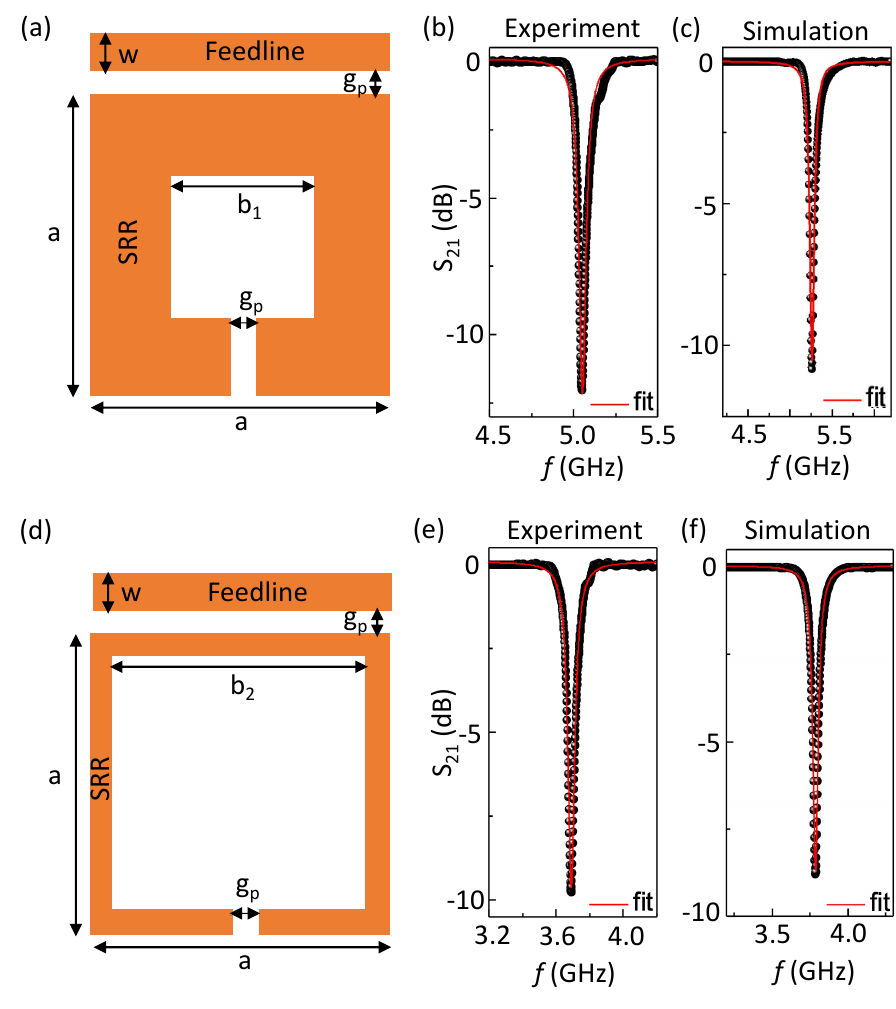}
\caption{(a), (d) Top view of two split-ring resonators used in the experiment with the dimensions defined in the text. (b), (e) Resonator modes of the two SRRs with the corresponding Lorentzian fit (red line). The resonating frequency of the SRR decreases from $\sim$ 5.05 GHz to $\sim$ 3.7 GHz by changing the inner width of the SRR from b$_1$ = 1.5 mm to b$_2$ = 2.5 mm. (c), (f) Ansys HFSS simulations of two resonators with the corresponding Lorentzian fits (red lines). The Ansys HFSS simulation gives resonating modes of ($5.261\pm0.002$) GHz and ($3.786\pm0.004$) GHz for the two designs, respectively.}
\label{srr_size}
\end{figure}

A schematic illustration of the experimental setup is shown in Fig.~\ref{schematic}. The magnonic hybrid system comprises an SRR loaded with epitaxial YIG spheres of varying diameters between 0.2 mm and 1 mm. 
The resonator mode with a dissipation rate  {$\kappa_\mathrm{p}/2\pi$} couples with the YIG magnon mode with a coupling constant $g_\mathrm{eff}~(=g/{2\pi})$, while the YIG sample dissipates its energy at a rate  {$\kappa_\mathrm{m}/2\pi$} as schematically illustrated in Fig.~\ref{schematic}(a).
The experiment requires a high-quality resonator to confine electromagnetic waves by reflecting them back and forth between the boundaries. The SRR is located in the vicinity of a microwave (MW) feedline, as shown in Fig.~\ref{schematic}(b). Ansys HFSS finite element simulations were used to optimize the SRR dimensions for the desired frequency. 
To perform the simulation, we set the SRR geometries, assigned the material properties to each SRR component, and defined the wave ports and boundary conditions. 
Based on the simulation results, we fabricated two SRRs with different resonance frequencies and, hence, different inner dimensions by etching one side of Rogers RO3010 laminate with a dielectric constant of 10.20 $\pm$ 0.30 and copper thickness of  {17.5~$\mu$m} that is coated on  {each side} of the substrate. 
The first SRR has the following dimensions:  outer and inner widths are $a = 4.5$ mm and $b_1 = 1.5$ mm, respectively, while the gap between the SRR and the feedline is $g_p = 0.2$ mm, and the feedline’s width is $w = 0.4$ mm, see Fig.~\ref{srr_size}(a). This SRR resonates at $\sim$ 5.05 GHz [see Fig.~\ref{srr_size}(c)] and has a quality factor, $Q_1=99.1\pm0.8$. The experimentally observed and Ansys HFSS simulated SRR resonances are shown in Figs.~\ref{srr_size}(b) and (c), respectively, with the corresponding fittings which are in good agreement  {with one another}: the values for simulated and fabricated SRR lie within 4\% of one another. Increasing the inner width of the SRR to $b_2$ = {2.5 mm} while keeping all other parameters the same decreases the resonating frequency to $\sim$ 3.7 GHz; see Fig.~\ref{srr_size}(d). This SRR has a slightly lower quality factor, $Q_2=74.8\pm0.5$. Figures~\ref{srr_size}(e) and (f) show the experimental and Ansys HFSS simulated SRR resonances for the larger inner ring. The corresponding fitting of simulated and fabricated SRR are in good agreement; the values lie within 2.5\% of one another. 

Microwave spectroscopy measurements were performed in the frequency domain using a vector network analyzer (VNA). The microwave signal was applied to one end of the feedline, and the transmission coefficient (S$_{21}$) was measured at the receiving port as a function of MW frequency and applied biasing field. The nominal microwave power was -15 dBm. The biasing field was applied in the plane of the resonator and perpendicular to the feedline, as shown in Fig.~\ref{schematic}(b).


 Ferrimagnetic insulator YIG (Y$_3$Fe$_5$O$_{12}$) single crystal spheres of different diameters were used as a magnon source.  The spherical geometry of the sample rules out an inhomogeneous demagnetization field~\cite{ivan18}. 
  We studied spheres of different diameters, i.e., the number of spins, in a planar resonator to test the coupling strength dependence on the number of spins. Furthermore, we investigated the dependence of coupling strength on the rf-magnetic field distribution, $h_\mathrm{rf}$, by changing the position of the YIG sphere with respect to the SRR. 

 \section{Results}
\begin{figure}[t]
\centering
\includegraphics[width=0.75\columnwidth]{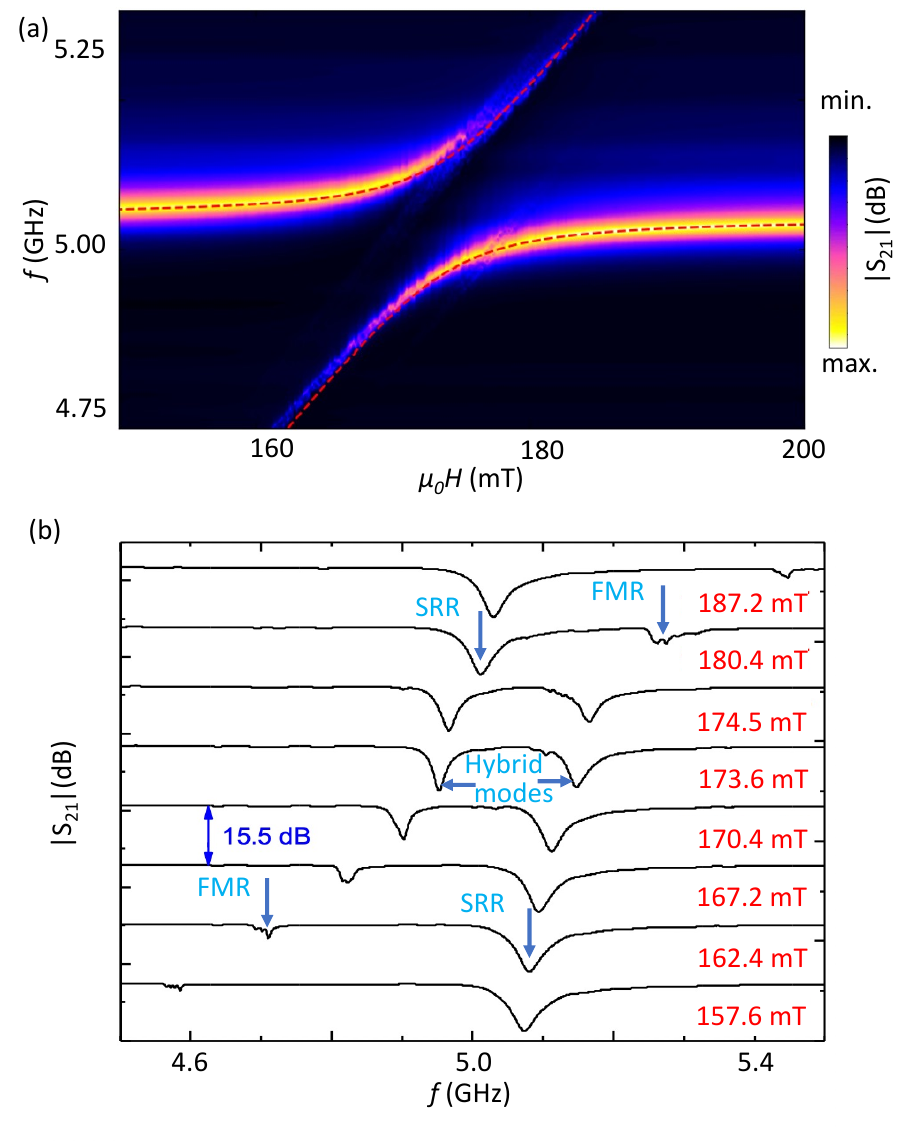}
\caption{ (a) A typical avoided level crossing spectrum where magnon mode  {and} SRR mode strongly couple. The spectrum was obtained using a YIG sphere (of radius 0.75 mm) placed in the center of the ring. The false color represents the S$_{21}$ transmission parameter. The dotted red curve is the fit based on Eq.~(\ref{MPfit}). (b) Transmission parameter S$_{21}$ as a function of frequency at different biasing magnetic field magnitudes.
}
\label{avoided crossing}
\end{figure}

 A typical room temperature avoided level crossing spectrum for a YIG sphere of radius 0.75 mm and SRR with $\sim$ 5.05 GHz resonance frequency is shown in Fig.~\ref{avoided crossing}. Figure~\ref{avoided crossing}(a) shows a false color-coded microwave absorption spectrum of the magnon-photon coupling where the color represents the transmission parameter (S$_{21}$). We observe an avoided crossing in the field/frequency region where we expect a crossing between the uncoupled magnon mode (magnetic-field dependent mode) and the microwave photon mode (horizontal, field-independent mode).  This avoided crossing corresponds to the hybridization between the two modes, creating two new dynamic hybrid modes, indicating the formation of a magnon polariton. The red dotted line in Fig.~\ref{avoided crossing}(a) is a fit to Eq.~(\ref{MPfit}) considering the ferromagnetic resonance condition for a spherical sample $f_\mathrm{r} = \gamma'H$, where $\gamma'$ is the reduced gyromagnetic ratio, and $H$ is the effective magnetic field. The parameters extracted are $\gamma' = 2.907\pm0.002$ GHz/kOe, which is close to the values reported for YIG spheres \cite{ferrisphere} and $g_\mathrm{eff}= 96.5\pm1.5$ MHz, which is larger than the value measured in a 3D-cavity for the same size of the sphere at low temperature~\cite{tabu14}. 
  {The cooperativity C relates the coupling rate ($g_\mathrm{eff}$) to the losses ($\kappa_\mathrm{p}/2\pi$ and $\kappa_\mathrm{m}/2\pi$): $\mathrm{C} = {g_\mathrm{eff}^2}/{(\kappa_\mathrm{p}/2\pi)(\kappa_\mathrm{m}/2\pi)}$}. 
For our system, the coupling strength $g_\mathrm{eff}$~(= 96.5 MHz) is larger than the losses $\kappa_\mathrm{p}/{2\pi}$~($=51$ MHz) and $\kappa_\mathrm{m}/{2\pi}$~($=20$ MHz). Hence, $\mathrm{C}$ $>$ 1, which means that the coupling lies in the strong coupling regime.

 We show exemplary frequency-dependent line plots of the transmission parameter (S$_{21}$) in Fig.~\ref{avoided crossing}(b) at different magnetic field magnitudes. When the field is swept from higher to lower values, the FMR mode approaches the SRR mode in frequency, while its intensity (S$_{21}$) increases. At $\sim$ 173.6 mT, the frequency gap is minimum and the two modes switch their intensities. The transduction between the magnon and photon modes is most efficient in this hybridized state. As the field is lowered further, the separation between the modes increases, with the lower-frequency mode -- the FMR mode -- having a lower intensity than the higher-frequency mode -- the SRR mode.

\begin{figure}[h!]
\centering
\includegraphics[width=0.75\columnwidth]{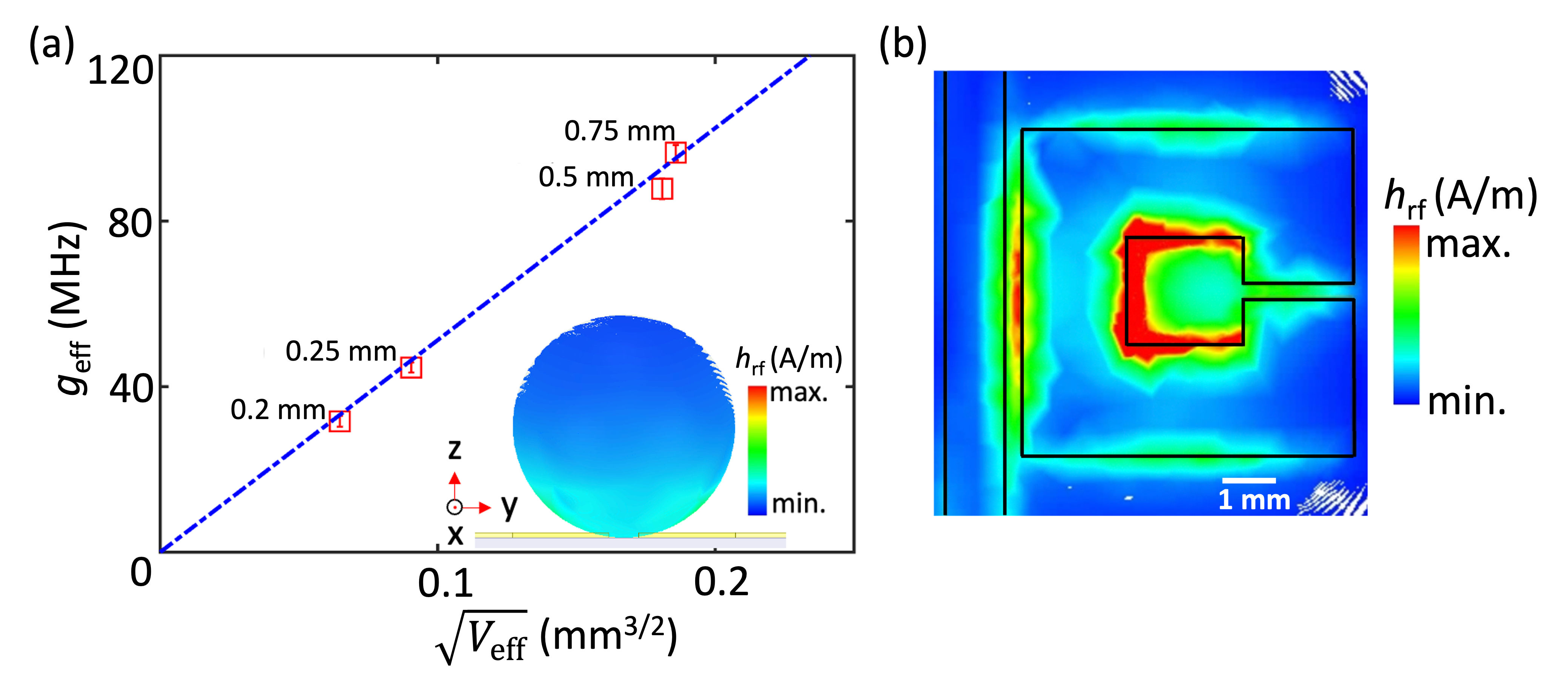}
\caption{(a) Coherent coupling strength as a function of the square root of the effective volume of the YIG sphere ($\mathrm{V_{a}}$) in an SRR (of frequency $\sim$ 5.05 GHz). The inset shows the $h_\mathrm{rf}$ distribution over the volume of the sphere with a diameter of 0.75 mm and (b) rf magnetic field distribution at 5.05~GHz obtained by Ansys HFSS simulations. 
}
\label{volume}
\end{figure}


To investigate the dependence of the coupling strength on the number of spins involved in the coupling, we performed the same type of experiment for different YIG sphere diameters. For this purpose, the YIG spheres were placed in the resonator's center. Figure~\ref{volume}(a) shows the variation of the coherent coupling strength with the square root of the \textit{effective} volume of the YIG sphere. {The effective volume is defined as the active material participating in the coupling. The effective volume was determined by calculating the microwave magnetic field ($h_\mathrm{rf}$) distribution using Ansys HFSS; see Fig.~\ref{volume}(b). The inset shown in Fig.~\ref{volume}(a) is the volume integral of the $h_\mathrm{rf}$ field for the 0.75 mm diameter sphere. As is evident from the figure, we can see that approximately one-fourth of the entire volume (the lower portion of the sphere) experiences a stronger $h_\mathrm{rf}$. Based on these modeling results, we define this one-fourth of the entire volume as the effective volume}. Figure~\ref{volume}(a) shows that the variation of coherent coupling strength on the square root of the effective volume is linear, following the predicted $\sqrt{N}$ dependence since the effective volume of the sphere $V_\mathrm{eff} \propto {N}$; $N$ - number of spins. Hence, the larger the YIG sample, the larger the effective volume, and the stronger the coupling, providing a control method of the coupling strength. 

\begin{figure}[h!]
\centering
\includegraphics[width=0.75\columnwidth]{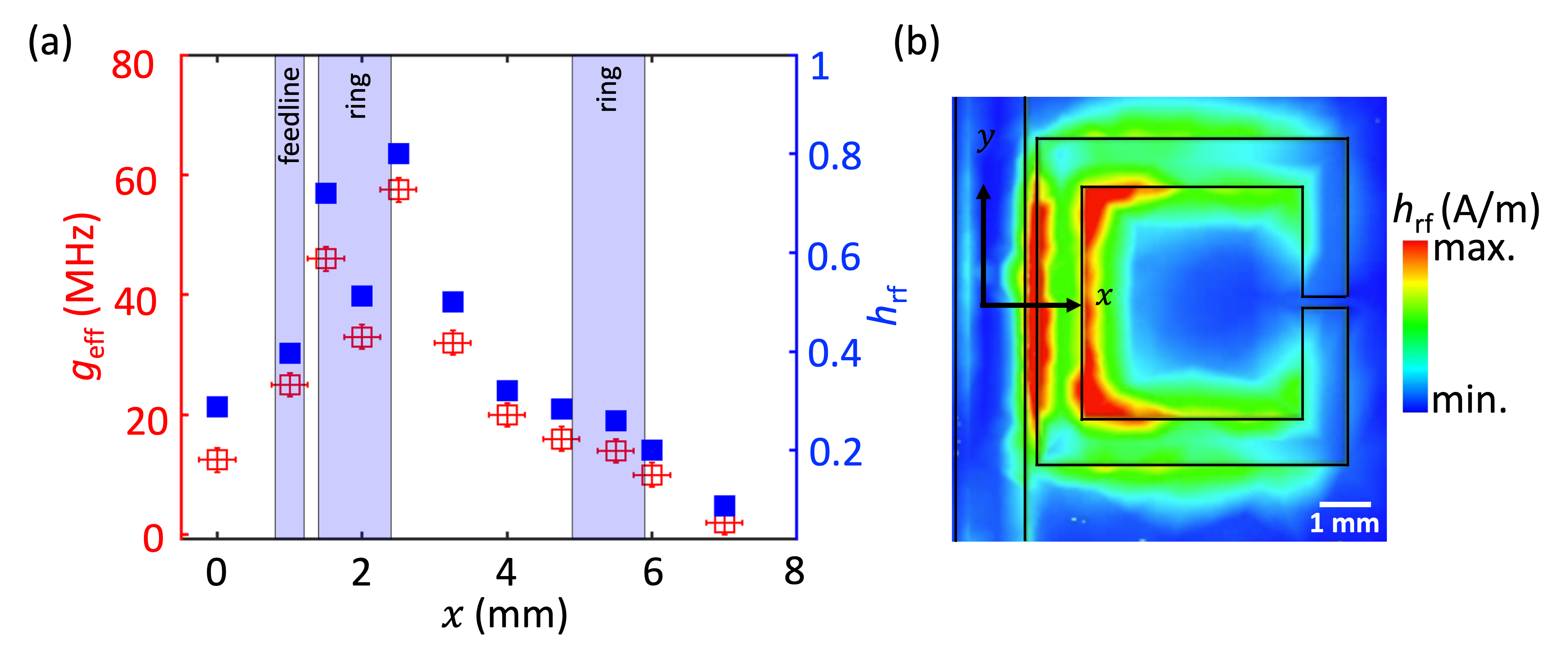}
\caption{(a) Coherent coupling strength (red) and normalized $h_\mathrm{rf}$ intensity (blue) as a function of position ($x$) of YIG sphere (of diameter 0.75 mm) in an SRR (of frequency $\sim$ 3.7 GHz). The shaded areas represent the positions of the feedline and the ring boundaries. (b) Rf magnetic field distribution obtained by Ansys HFSS simulation for the 3.7~GHz SRR. The color scales in Fig.~\ref{position}(b) and Fig.~\ref{volume}(b) are normalized to the same value, so they are directly comparable.}
\label{position}
\end{figure}

Furthermore, we investigated the position dependence of the magnon-photon coupling along the $x$-axis of the SRR (as is shown in Fig. ~\ref{position}). This position-dependent experiment was conducted using a YIG sphere with a diameter of 0.75 mm and an SRR with a resonance frequency of $\sim$ 3.7 GHz. Figure~\ref{position}(a) compares the experimentally observed $g_\mathrm{eff}$-dependence (red data points, left axis) with the rf magnetic field ($h_\mathrm{rf}$) distribution obtained by Ansys HFSS simulation for an SRR with resonance frequency at $\sim$ 3.7 GHz (blue data point, right axis). The simulated position-dependent rf magnetic field distribution along the $x$-axis was extracted from the two-dimensional field distribution shown in Fig.~\ref{position}(b). Our results reveal that the coherent coupling strength $g_\mathrm{eff}$ varies non-monotonically [see Fig.~\ref{position}(a)]. Furthermore, we find a direct correlation between the coupling strength and the magnitude of the microwave magnetic field. This indicates a stronger mode overlap between the spins in the YIG sphere and the magnetic component of the electromagnetic wave created by the SRR photon when the rf magnetic field $h_\mathrm{rf}$ is stronger. The maximum coupling strength ($57$~MHz) is observed close to the inner wall of the SRR-ring where  $h_\mathrm{rf}$ is maximum [see Fig.~\ref{position}(a)]. As we move further away from this position, the coupling strength decreases. In other words, the weaker the magnetic field generated by the SRR mode, the smaller will be the coupling strength, enabling an active control of the coupling strength by varying the spatial location of the YIG sphere on the SRR. We also note that the observed correlation of the magnon-photon coupling strength with the created rf magnetic field could be utilized as a magnetic field sensor in a properly calibrated system.

When comparing the rf magnetic field distributions for the two resonator designs [Figs.~\ref{volume} (b) and ~\ref{position}(b)], it becomes clear that the modeled microwave magnetic field strength $h_\mathrm{rf}$ in the center of the ring is greater for the $5.05$ SRR than for the $3.7$~GHz SRR. This means a stronger coupling of a particular YIG sphere with a given diameter can be achieved with the $5.05$ GHz SRR compared to the $3.7$ GHz SRR [Fig.~\ref{position}(b)]. We extracted a coupling strength of $g_{5.05}\approx96$ MHz [see experimental results in Fig.~\ref{avoided crossing}(a)] for a 0.75 diameter sphere coupled to the $5.05$ GHz, while a coupling strength of $g_{3.7}\approx20$ MHz was found for the $3.7$ GHz resonator [see Fig.~\ref{position}(a) for $x$=4~mm, which is approximately in the center of the SRR].
In the following, we determine the expected ratio between the coupling strengths of the two resonators for a given YIG sphere diameter (0.75 mm). The coupling strength is given by $g_\mathrm{eff}\propto \sqrt{\omega V_m/V_a}$, where $\omega $ is the resonator frequency, $V_a$ and $V_m$ are the mode volume of the resonator and YIG sphere volume, respectively~\cite{zhang14}. Here, the resonator mode volume can be approximated by the volume integral of the $h_{rf}$ field distribution over the volume of the inner space of the resonator; the dependence on $V_\mathrm{m}$ cancels out when we calculate the ratio of the coupling strengths since the same YIG sphere diameter was used. With this approximation, one can express the ratio of $g_\mathrm{eff}$ for two resonators as $g_\mathrm{5.05}/g_\mathrm{3.7}=3.86$, which is close to the ratio extracted from the experiment ($g_\mathrm{5.05}/g_\mathrm{3.7}\approx{96/20} = 4.8$). 

We note that Shi et al.~\cite{shi_JPD_19} observed a different behavior: when the YIG sample is placed in the highly non-uniform microwave magnetic field region near the edge of the SRR, a larger coupling was observed for the lower-frequency resonator. To confirm if we find the same behavior, we tested the coupling strength when the YIG sphere was placed at the edge of the resonators facing the feedline (not in the center of the ring as discussed above) and found an agreement with Shi et al.'s observation: the coupling strength for the 5.05 GHz SRR was 15 MHz when the sample is placed close to the SRR edge, while a larger coupling strength of ~$37$~MHz was found for the 3.7 GHz SRR at the same position [see also position dependence in Fig.~\ref{position}(a)]. Finally, we note that this behavior can be expected based on the correlation between the coupling strength and the microwave magnetic field discussed above: $h_\mathrm{rf}$ has a non-uniform position dependence that is maximum near the edge of the SRR [see Fig.~\ref{position}(a)].

\section{Outlook}

In conclusion, we demonstrated the coherent coupling of magnons with the microwave magnetic field in an SRR/YIG sphere hybrid system at room temperature. Our results show that the coherent coupling can be controlled by 
varying the YIG sphere diameter and, hence, the number of spins participating in the coupling 
modifying the mode overlap through dipolar (Zeeman) interaction between the spins in the YIG sphere with the magnetic component of the electromagnetic wave of the SRR photon mode. By comparing two distinct SRRs with different resonances, we furthermore confirmed the theoretically expected dependence of the coupling strength on both the resonating frequency and the effective mode volume of the resonator.

\section{Acknowledgment}
We acknowledge support by the U.S. Department of Energy, Office of Basic Energy Sciences, Division of Materials Sciences and Engineering under Award DE-SC0020308.

\section{References}
\bibliographystyle{iopart-num} 

\end{document}